# Valley Polarization Enhancement Induced by a Single Chiral Nanoparticle

Sejeong Kim,[1,†,*] Yae Chan Lim,[2,†] Ryeong Myeong Kim,[2] Johannes E. Fröch,[1] Thinh N. Tran,[1] Ki Tae Nam[2,*] and Igor Aharonovich[1,3]

[1]School of Mathematical and Physical Sciences, University of Technology Sydney, Ultimo 2007, New South Wales, Australia

[2]Department of Materials Science and Engineering, Seoul National University, Seoul, 08826, Republic of Korea

[3]ARC Centre of Excellence for Transformative Meta-Optical Systems, Faculty of Science, University of Technology Sydney, New South Wales, Ultimo, 2007, Australia

[†]These authors contributed equally.

[*]Sejeong.Kim-1@uts.edu.au , nkitae@snu.ac.kr

**Abstract:** Valley polarization is amongst the most critical attributes of atomically thin materials. However, achieving a high contrast from monolayer transition metal dichalcogenides (TMDs) has so far been challenging. In this work, a giant valley polarization contrast up to 45% from a monolayer $WS_2$ has been achieved at room temperature by using a single chiral plasmonic nanoparticle. The increased contrast is attributed to the selective enhancement of both the excitation and the emission rate having one particular handedness of the circular polarization. The experimental results were corroborated by the optical simulation using finite-difference time-domain (FDTD) method. Additionally, the single chiral nanoparticle enabled the observation of valley-polarized luminescence with a linear excitation. Our results provide a promising pathway to enhance valley contrast from monolayer TMDs and utilize them for nanophotonic devices.

Transition metal dichalcogenides (TMDs) have become vital enablers to explore unique nanophotonic and optoelectronic phenomena.[1-3] A key example is a valleytronics, which harnesses the valley degree of freedom in semiconductors.[4-7] The broken inversion symmetry of TMDs enables to selectively excite and optically readout the valley degree of freedom using circularly polarized light.[8, 9] It is crucial to increase the contrast of valley-dependent photoluminescence (PL) to build practical valleytronics devices for quantum information processing.[10-13] Yet, obtaining a clear valley contrast at room temperature (RT) is challenging, due to its severe contrast reduction by phonon-assisted scattering processes. Consequently, the majority of the valley polarization experiments are limited to cryogenic temperature.

To circumvent this problem, various strategies have emerged so far including deploying multilayer configurations.[10, 14] In such configuration, the main obstacle that reduces the valley contrast at room temperature, *i.e.* phonon-assisted hole scattering through Γ valley, is significantly suppressed and the valley polarization contrast up to 77% is demonstrated using 4-layered $WS_2$.[14] However, such schemes suffer from an extremely low luminescence due to indirect bandgap transition from multilayer TMDs, therefore, achieving a large valley contrast with monolayer TMDs is a crucial milestone. The utilization of chiral nanostructured surfaces is suggested as promising candidates, which include plasmonic and dielectric chiral metasurface arrays as well as gold moiré chiral patterns.[12, 15-17] These chiral nanostructures could efficiently enhance the valley contrast up to 43% at room temperature with monolayer,[18] yet the attempts so far required cumbersome and sophisticated nanofabrication protocols to make nanostructured arrays.

Bottom-up fabrication approaches provide powerful routes to address complex nanofabrication issues. Chemically synthesized chiral gold nanoparticles are recently achieved by the simple solution-based synthesis method creating a uniform and large amounts of chiral nanostructures.[19, 20] Among diverse chemically-synthesized chiral gold nanoparticles, 432 helicoid III show an exceptional chiroptical response, showing Kuhn's dis-symmetry factor (g-factor, $g = 2\frac{I_{LHC}-I_{RHC}}{I_{LHC}+I_{RHC}}$) of 0.2 when the chiral gold nanoparticles are in an ensemble. The g-factor further increases up to 0.8, which is one of the highest value among chiral metamaterials.[21] Such strong chiroptical response opens a pathway of creating single-nanoparticle controlled nanophotonic systems.

Here we realize a facile, single-step technique using the chiral gold nanoparticles to dramatically enhance the valley polarization, achieving a contrast of up to 45% at RT. Interestingly, the valley contrast is enhanced by the effect of a single nanoparticle that increased the degree of valley-polarization from 16% to 45%. Photonic simulation based on the finite-domain time-difference (FDTD) method reveals the selective enhancement of the excitation and the emission rates by the chiral gold nanoparticle when the light is right-handed circularly polarized. The method presented in this work is scalable and does not require sophisticated nanofabrication protocols; hence, offers a promising pathway to be used with any 2D materials.

**RESULTS AND DISCUSSION**

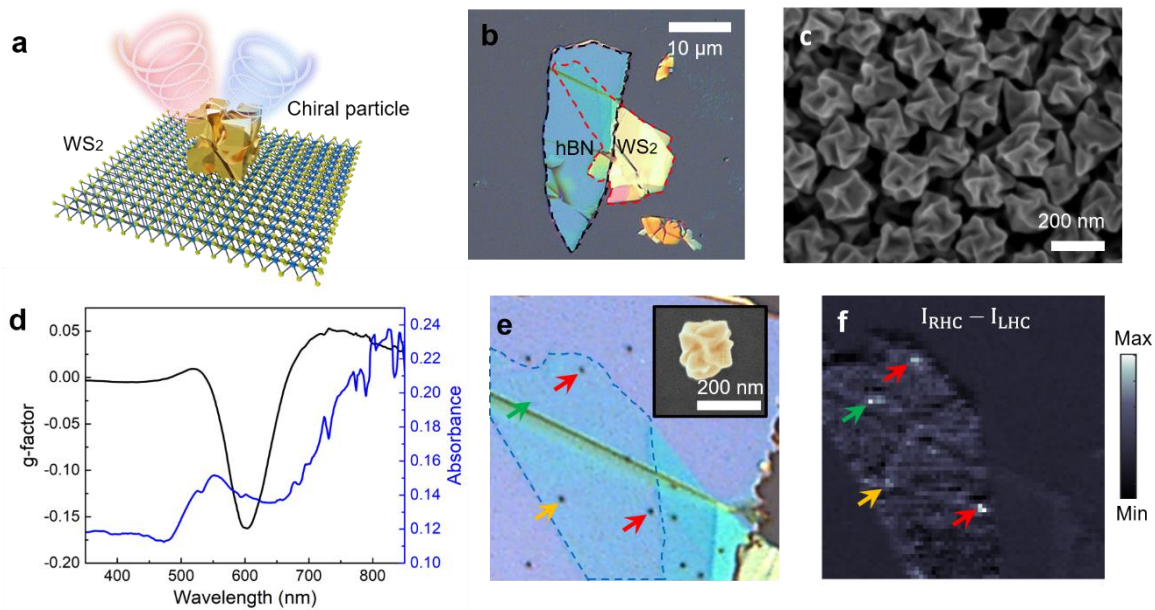

**Figure 1.** (a) 3D schematic showing a chiral gold nanoparticle on $WS_2$ under left and right circularly polarized illumination. (b) An optical microscope (OM) image of the sample, consisting of a $WS_2$ monolayer transferred onto an hBN flake. (c) A SEM image of chemically synthesized chiral gold nanoparticles named 432 helicoid III. (d) Experimental g-factor (black line) & absorption (blue line) spectra of 432 helicoid III nanoparticles. (e)

An OM image showing the chiral gold nanoparticles (black dots) on top of the 2D stack. Inset is a false-color SEM image of the chiral gold nanoparticle. (f) A PL intensity difference map showing enhanced contrast, corresponding to the position of nanoparticles on the $WS_2$ monolayer.

Figure 1a shows the schematic illustration of our system, consisting of a single chiral gold nanoparticle positioned on top of a monolayer $WS_2$. The chiral gold nanoparticle preferably absorbs the right-handed circularly polarized light and shows stronger plasmonic effects. The mechanically exfoliated $WS_2$ (red dashed line) is first transferred onto hBN (black dashed line) for better exciton quality,[22] as shown in Figure 1b. A PVA stamp is used to pick-up and release the 2D materials. The chiral gold nanoparticles are then drop-casted onto the 2D stack. These chiral gold nanoparticles (432 helicoid III, see Method section) are chemically synthesized using seed-mediated growth of 40 nm octahedron gold seed nanoparticles with amino acid and peptides as shape modifying molecules to create helicoidal morphology.[19, 23] This water-based, bottom-up fabrication approach enables them to produce large and uniform quantities of 150 nm chiral gold nanoparticles with remarkable chiroptical properties, as shown in Figure 1c. Figure 1d shows a measured g-factor and absorption spectra of 432 helicoid III nanoparticles. The result shows a strong resonance at $\lambda=600$ nm, which implies selective plasmonic resonance occurring at a particular wavelength with right-handed circularly polarized (RHC) incident light. A g-factor of -0.16 (0.06) is observed at $\lambda=600$ nm (750 nm) in an ensemble, which indicates the nanoparticles preferably absorb R(L)HC light. For absorption and CD measurement, chiral gold nanoparticles were prepared in a solution state, dispersed in CTAB (1 mM).

A magnified optical image of the final device is shown in Figure 1e, with several nanoparticles clearly visible on the monolayer region with blue dashed lines. Each black dot is a single chiral nanoparticle as shown in the inset SEM image. Next, photoluminescence (PL) intensity from the sample is measured with 594 nm pulsed laser (40 MHz, 0.2 µW) scanning through the sample. The 100 **x** objective lens with a numerical aperture of 0.9 is used to focus the excitation laser as well as to collect PL signals. The excitation laser is circularly polarized by using a combination of a linear polarizer and a quarter-wave plate. Figure 1f shows the PL intensity difference where bright spots (red arrows) correspond to the location of chiral gold nanoparticles, confirming that the single chiral gold nanoparticle enhances the PL from the $WS_2$. While a single nanoparticle effect on circularly polarized light dichroism is clearly observed, a bright spot pointed by a green arrow in Figure 1f is also observed which might be induced by the wrinkles in 2D materials or the residual polymer on $WS_2$ accumulated during the transfer process. We also note that chiral gold nanoparticles are not entirely identical,[21] which caused the relatively low circular dichroism indicated by yellow arrows in Figures 1e, f. However, uniformity of such chiral gold nanoparticles can be significantly increased by adopting multi-chirality-evolution step synthesis method.[24]

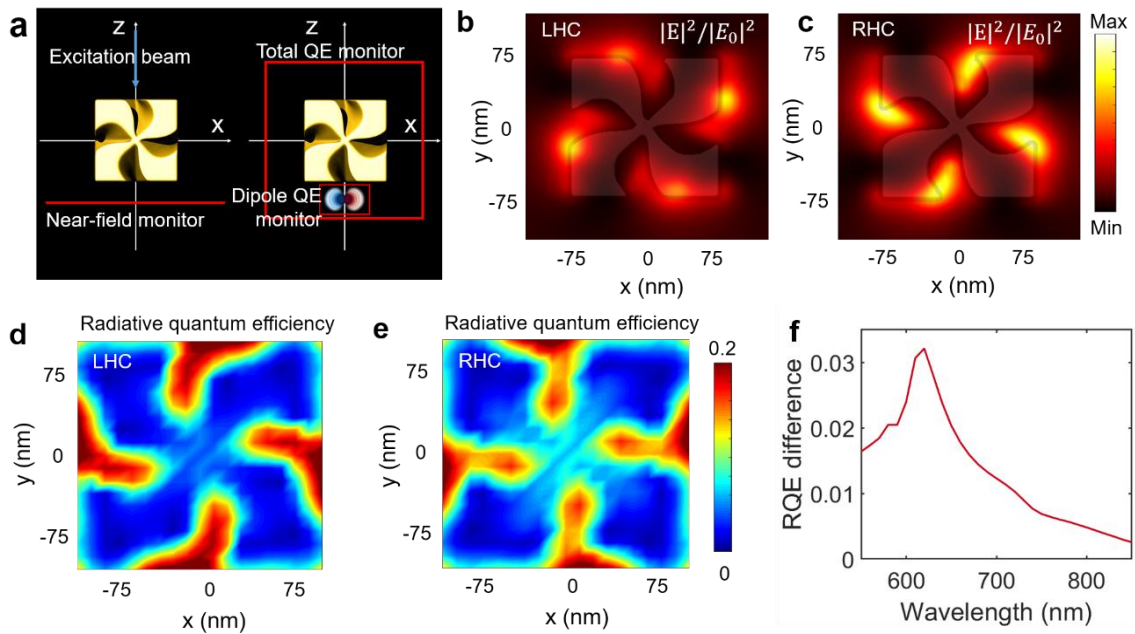

**Figure 2.** (a) Schematics of photonic simulation set-ups for the calculation of near-field intensity (left) and radiative quantum efficiency (right). Calculated near-field distribution under the incidence of (b) LHC and (c) RHC light at 600 nm. Maps of the radiative quantum efficiencies relative to a dipole emitter in free space under (d) LHC and (e) RHC emission at 620 nm, and (f) their difference depending on the wavelength were obtained.

The chiral gold nanoparticle plays two roles to enhance the valley polarization in a given configuration. First, the excitation light with RHC induces collective oscillation of localized surface plasmons in chiral gold nanoparticles, which enhances the electric field intensity in the near-field. Second, the same particle also enhances the PL emission with RHC that carries the valley information. The large valley contrast enhancement is the combination of both these effects. These two effects are further analyzed by using the finite-difference time-domain (FDTD) method (Lumerical FDTD solver), and the simulation set-up is depicted in Figure 2a. For the near-field intensity calculation, the monitor was located beneath the chiral gold nanoparticle, *i.e.* where $WS_2$ layer is located, under the LHC and RHC illumination, respectively. In the case of calculating the radiative quantum efficiency (RQE), two enclosed monitors are used: a total QE monitor that measures the total power radiated from the system consists of a dipole source and the chiral particle, and a dipole QE monitor that measures the radiated power from the dipole source.

First, the near-field intensity profiles of the chiral gold nanoparticle are shown under the LHC (Figure 2b) and RHC (Figure 2c) light incidence with λ=600 nm. The simulation results show the electric-field intensity is enhanced by 21% under RHC excitation compared to LHC excitation. Next, PL enhancement is analysed, which is a combination effect from the spontaneous rate enhancement and the radiative quantum efficiency (RQE) enhancement.[25] Due to the large field enhancement and increased density of states near the chiral gold nanoparticles, $WS_2$ experiences increased spontaneous emission rate. Besides, RQE of emission plays a role determining PL enhancement. To calculate the RQE, the radiative decay rate of the dipole, the non-radiative decay rate from photon absorption by the particle, and the total radiative decay rate of the system are considered. The spatial dependence of the RQE with respect to the position of the emitter of the system under LHC and RHC emission is displayed in Figure 2d, e, respectively. Moreover, Figure 2f shows the difference in RQE, *i.e.* (RQE with RHC) - (RQE with LHC), which clearly indicates RQE enhancement at the chiral gold nanoparticle's resonance. This result shows the PL emission is also selectively enhanced depending on the handedness of the emitted light.

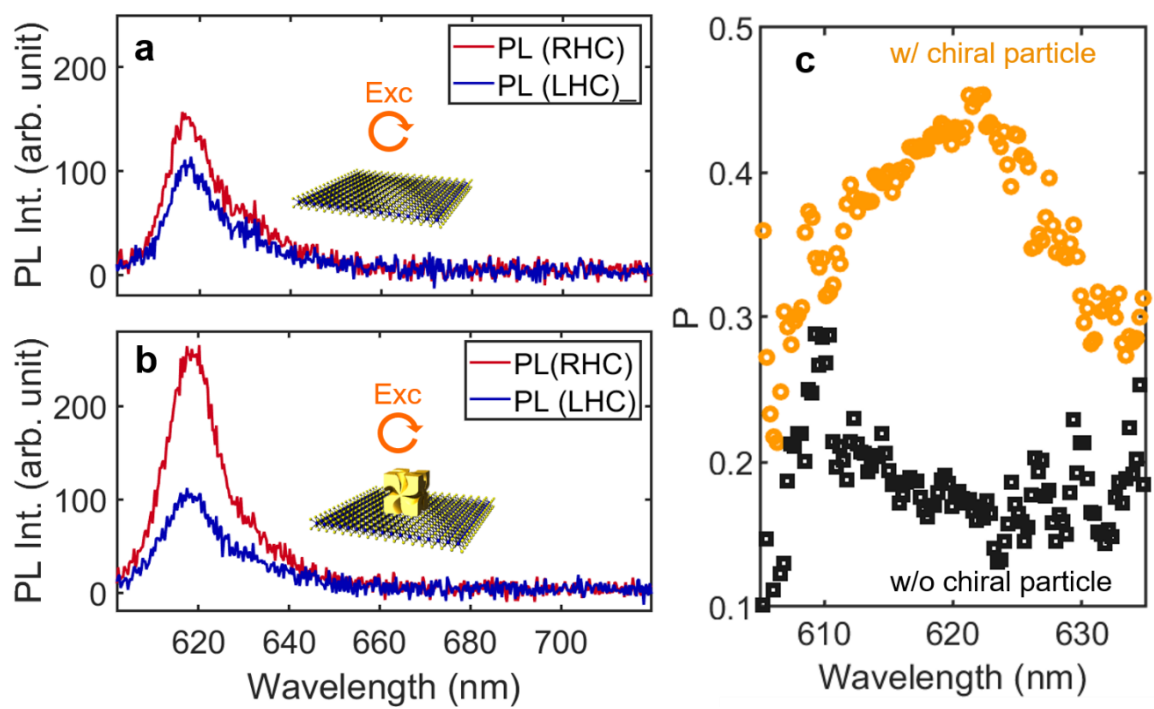

**Figure 3.** Polarization resolved PL emission from (a) WS$_2$ only and (b) with a chiral particle on WS$_2$ when the excitation laser is RHC light. (c) Degree of polarization (P) plot of WS$_2$ with (orange circles) and without (black circles) a chiral gold nanoparticle on top.

To quantitatively confirm the degree of valley contrast induced by the chiral gold nanoparticles, we compare pristine WS$_2$ samples to those with nanoparticles on top. All measurements were carried out under ambient condition at room temperature. A yellow laser (594 nm, pulsed 40 MHz) is used for near-resonant excitation. An objective lens with NA=0.9 is used to excite the sample and collect the signal, where the pump spot size of the excitation beam is approximately 500 nm in diameter. The RHC polarized laser excites predominantly the K valley, which results in a higher RHC PL intensity, as shown in Figure 3a. The valley-polarization contrast at room temperature is typically small or negligible due to the spin-flip and intervalley scattering. Next, the polarization-dependent spectra were measured from the WS$_2$ monolayers integrated with the chiral gold nanoparticles, as shown in Figure 3b. Notably, the valley polarization is dramatically enhanced by a single chiral gold nanoparticle showing the strong selective response of the chiral gold nanoparticle to circularly polarized light. The degree of polarization, which is defined as $P =$

$\frac{PL\ Int(RHC) - PL\ Int\ (LHC)}{PL\ Int(RHC) + PL\ Int\ (LHC)}$, is extracted from Figures 3a, b and plotted in Figure 3c. The graph shows the clearly enhanced valley polarization by the single chiral gold nanoparticle where the maximum degree of polarization achieves 0.45.

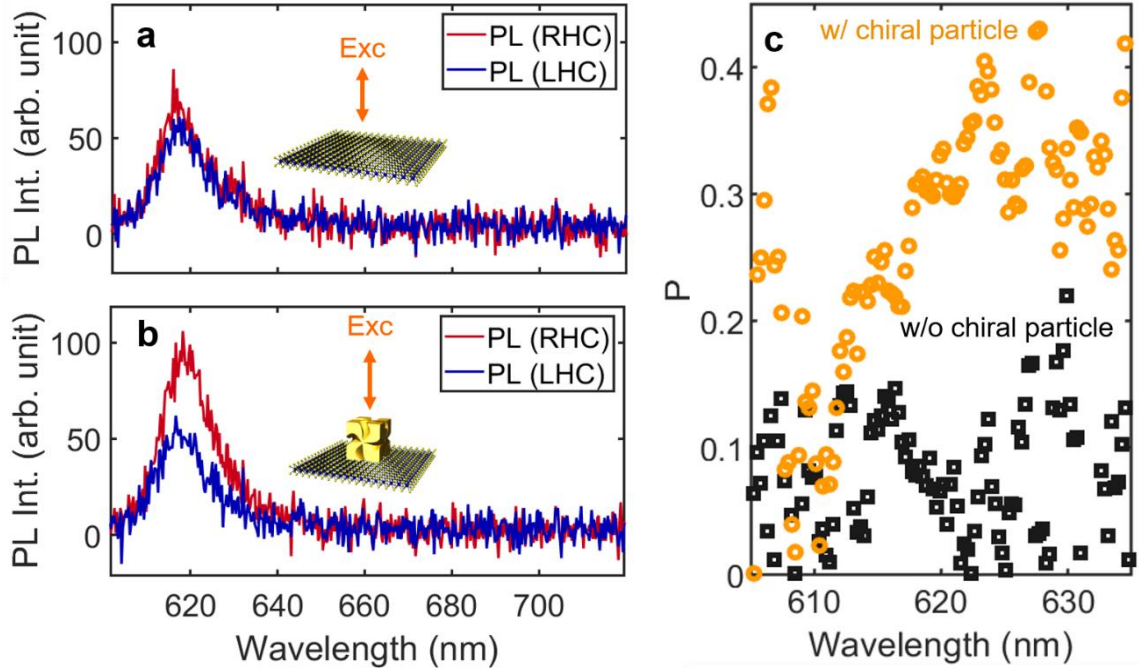

**Figure 4.** Polarization resolved PL emission from (a) WS$_2$ only and (b) with a chiral particle on WS$_2$ when the excitation laser is linearly polarized. (c) Degree of polarization (P) plot of WS$_2$ w/ and w/o a chiral gold nanoparticle.

Furthermore, valley-polarization dependence with linear polarization excitation is investigated. It is expected that the PL intensity from RHC and LHC are equal as linearly polarized light excites both K and K′ valleys. Figure 4a shows the comparable PL intensity with polarization contrast of 0.06. This value is considered to be the system error which is a combined effect from optics used in the setup. The WS$_2$ monolayer-chiral gold nanoparticle configuration further provides the ability to distinguish valley polarization with linearly polarized excitation light, as shown in Figure 4b. Excitation with linear light is advantageous for fast screening of valley devices based on 2D materials. Degree of polarization is plotted in Figure 4c showing the valley contrast over 0.3 with linearly polarized excitation.

On-chip devices transporting valley information at room temperature is desirable for nanophotonics. Our design of placing a chemically-synthesized chiral gold nanoparticle on 2D material immediately solves one of the challenging problems in creating chiral surface- 2D monolayer configuration. Transferring TMDs on top of the chiral metasurfaces causes strain-induced intervalley scattering,[26] which degrades valley polarization. On the other hand, depositing metals on TMDs also found to have a detrimental effect on the material quality.[27] Therefore, employing chemically synthesized chiral gold nanoparticles suggest a robust tool for studying chiral plasmonics assisted valleytronics.

## CONCLUSION

To conclude, we report on record enhancement of valley polarization contrast of ~ 45% at room temperature from a monolayer $WS_2$. This is achieved by positioning a single, purposely engineered chiral gold nanoparticle on top of the monolayer. A single chiral nanoparticle enhanced valley-dependent PL contrast of a monolayer WS2 from 16% to 45%, benefited by the large chiroptical response of a chemically synthesized nanoparticle. The bottom-up, water-based fabrication method provides a way to place a chiral nanostructure without damaging the TMDs. We conducted a photonic simulation based on FDTD methods to confirm the asymmetric optical response of the chiral gold nanoparticle. Our approach is scalable and can be utilized for a plethora of 2D materials, as well as to realize spatially selective valley enhancement in nanophotonic devices.

**Methods**

**Chemicals.** Hexadecyltrimethylammonium bromide (CTAB, 99%) was purchased from Alfa Aesar. L-ascorbic acid (AA, 99%), L-glutathione (L-GSH, 98%) and tetrachloroauric(III) trihydrate ($HAuCl_4 \cdot 3H_2O$, 99.9%) were purchased from Sigma-Aldrich. All aqueous solutions were prepared using high-purity deionized water (18.2 MΩ cm$^{-1}$).

**Synthesis of 432 helicoid III nanoparticles.** 40 nm octahedron seed nanoparticles were synthesized referring to a previously reported article.[28] Synthesized octahedron gold seed nanoparticles were centrifuged (6,708 g, 150 s) twice and washed with CTAB (1 mM) solution. After the washing process, nanoparticles were re-dispersed in CTAB (1 mM) solution. In a synthesis of 432 helicoid III nanoparticles, a growth solution was prepared by adding 0.8 ml of CTAB (100 mM) and 0.2 ml of $HAuCl_4$ (10 mM) into 3.95 ml of deionized water to form an $[AuBr]^-$ complex. $Au^{3+}$ was then reduced to $Au^+$ by the rapid injection of 0.475 ml of AA (100 mM) solution. The growth of chiral nanoparticles was initiated by injecting 5 μl of L-GSH (5 mM) followed by addition of 50 μl of octahedron gold seed nanoparticles into the growth solution. The growth solution was left undisturbed in a 30 °C bath for 2 h, and the purple solution gradually became blue with large scattering. After 2 h growth, the solution was centrifuged twice (1,677 g, 60 s) to remove unreacted reagents and was re-dispersed in CTAB (1 mM) solution for further use.

**Preparation of TMDC and TMDC-432 helicoid III nanoparticle complex.** $WS_2$/hBN heterostructure was prepared using a dry-transfer method with a poly(vinyl alcohol) (PVA) stamp.[29] Both $WS_2$ and hBN flakes were mechanically exfoliated onto polydimethylsiloxane (PDMS) films first, then transferred onto $SiO_2$ substrate using PVA as a stamp. The PVA stamp with the above heterostructure was then released onto a marked thermally oxide silicon substrate, and the PVA was dissolved in warm DI water for 3 hours. The PVA and 2D materials were heated to 60 °C during the pick-up steps and to 90 °C for the final release process. The helicoid nanoparticles were drop cast onto heterostructure.

**FDTD simulation analysis.** The chiral gold particle model was constructed using a 3D computer-aided design software (Rhinoceros 5.0), and its suitability was verified in the previous literature.[19] The near-field and the quantum efficiency calculation were conducted to examine the enhancement of excitation and emission by the chiral gold particle using a 3D vectorial Maxwell equation solver based on the finite-difference time-domain (FDTD) method (Lumerical FDTD 8.0).

**CD characterization.** Absorption and circular dichroism (CD) spectra were obtained using a J-815 spectropolarimeter instrument (JASCO). For absorption and CD measurement, chiral gold nanoparticle sample is prepared in a solution state (dispersed in CTAB (1 mM)).

**Photoluminescence measurement.** All the PL measurements were conducted at room temperature using a 594 nm laser as an excitation laser. An avalanche photodiode (APD) is used for PL mapping shown in Figure 1f, and a spectrometer (300 g/mm) is used for PL spectra in Figure 2 and Figure 4. The circularly polarized excitation light is created using a quarter waveplate and a linear polarizer while the emission from $WS_2$ is analyzed by additional quarter waveplate, half waveplate, and a linear polarizer.